\documentclass{sigchi}

\toappear{\scriptsize Permission to make digital or hard copies of all or part of this work for personal or classroom use is granted without fee provided that copies are not made or distributed for profit or commercial advantage and that copies bear this notice and the full citation on the first page. Copyrights for components of this work owned by others than ACM must be honored. Abstracting with credit is permitted. To copy otherwise, or republish, to post on servers or to redistribute to lists, requires prior specific permission and/or a fee. Request permissions from permissions@acm.org. \\
{\emph{CHI '20, April 25--30, 2020, Honolulu, HI, USA.} } \\
Copyright is held by the owner/author(s). Publication rights licensed to ACM. \\
ACM ISBN 978-1-4503-6708-0/20/04\ ...\$15.00.\\
http://dx.doi.org/10.1145/3313831.3376218}

\usepackage{balance}       
\usepackage{graphics}      
\usepackage[T1]{fontenc}   
\usepackage{txfonts}
\usepackage{mathptmx}
\usepackage[pdflang={en-US},pdftex,pdfpagelabels=false]{hyperref}
\usepackage{xcolor, colortbl}
\usepackage{multirow}
\usepackage{booktabs}
\usepackage{textcomp}
\usepackage{enumitem}
\usepackage{mathtools}
\usepackage{soul}
\usepackage{dialogue}
\usepackage{lips}
\usepackage{attrib}
\usepackage[caption=false]{subfig}
\usepackage{mfirstuc}
\usepackage{makecell}
\usepackage{arydshln}

\usepackage{microtype}        
\usepackage{ccicons}          

\def\plaintitle{ArguLens: Anatomy of Community Opinions On Usability Issues Using Argumentation Models}

\def\emptyauthor{}
\def\plainkeywords{Open source software; usability; online communities; issue discussion analysis; argumentation analysis.}


\makeatletter
\def\url@leostyle{%
  \@ifundefined{selectfont}{
    \def\UrlFont{\sf}
  }{
    \def\UrlFont{\small\bf\ttfamily}
  }}
\makeatother
\def\pprw{8.5in}
\def\pprh{11in}

\definecolor{linkColor}{RGB}{6,125,233}
\hypersetup{%
  pdftitle={\plaintitle},
  pdfauthor={\emptyauthor},
  pdfkeywords={\plainkeywords},
  pdfdisplaydoctitle=true, 
  bookmarksnumbered,
  pdfstartview={FitH},
  colorlinks,
  citecolor=black,
  filecolor=black,
  linkcolor=black,
  urlcolor=linkColor,
  breaklinks=true,
  hypertexnames=false
}

\definecolor{lightgray}{gray}{0.9}
\newcommand{\clgrey}[1]{{\cellcolor{lightgray}{#1}}}  

\begin{document}

\title{\plaintitle}

\numberofauthors{5}
\author{%
  \alignauthor{Wenting Wang\\
    \affaddr{McGill University}\\
    \affaddr{Montr\'eal, Canada}\\
    \email{wenting.wang@mail.mcgill.ca}}\\
  \alignauthor{Deeksha Arya\\
    \affaddr{McGill University}\\
    \affaddr{Montr\'eal, Canada}\\
    \email{deeksha.arya@mail.mcgill.ca}}\\
  \alignauthor{Nicole Novielli\\
    \affaddr{University of Bari}\\
    \affaddr{Bari, Italy}\\
    \email{nicole.novielli@uniba.it}}\\
  \alignauthor{Jinghui Cheng\\
    \affaddr{Polytechnique Montr\'eal}\\
    \affaddr{Montr\'eal, Canada}\\
    \email{jinghui.cheng@polymtl.ca}}\\
  \alignauthor{Jin L.C. Guo\\
    \affaddr{McGill University}\\
    \affaddr{Montr\'eal, Canada}\\
    \email{jguo@cs.mcgill.ca}}\\
}

\teaser{
\centering
\includegraphics[width=1\linewidth]{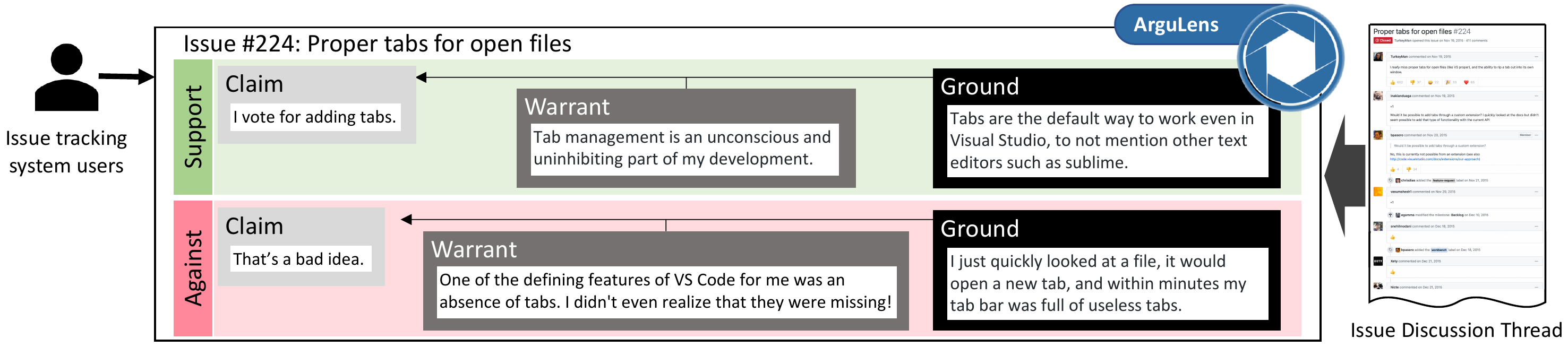}
\caption{ArgueLens analyzes lengthy discussions about usability on issue tracking systems for open source projects, anatomizing them into argumentative components (claim, ground, and warrant) and standpoints (support or against). Such a conceptual framework and the underlying argument extraction technique enable tools for supporting open source community members to understand and consolidate diverse opinions on usability issues.}
\label{fig:teaser}
}

\maketitle

\begin{abstract}
In open-source software (OSS), the design of usability is often influenced by the discussions among community members on platforms such as issue tracking systems (ITSs). However, digesting the rich information embedded in issue discussions can be a major challenge due to the vast number and diversity of the comments. We propose and evaluate ArguLens, a conceptual framework and automated technique leveraging an argumentation model to support effective understanding and consolidation of community opinions in ITSs. Through content analysis, we anatomized highly discussed usability issues from a large, active OSS project, into their argumentation components and standpoints. We then experimented with supervised machine learning techniques for automated argument extraction. Finally, through a study with experienced ITS users, we show that the information provided by ArguLens supported the digestion of usability-related opinions and facilitated the review of lengthy issues. ArguLens provides the direction of designing valuable tools for high-level reasoning and effective discussion about usability.


\end{abstract}



\section{Introduction}


Usability is a well-known, yet important concept that focuses on creating a system that is easy, efficient, error-preventing, and pleasant to be used~\cite{Nielsen1993}. In community-driven software development environments, such as that of the modern Open Source Software (OSS) applications, usability considerations are usually determined by the collective concerns of the community itself~\cite{Andreasen2006}, which, in turn, comprises a heterogeneous group of participants~\cite{cheng2019activity}. 

In the recent years, OSS applications vastly broadened their scope, going beyond the tools for ``technology adventurers'' while embracing a wide range of applications for diverse tasks and user groups. OSS development has since evolved beyond the ``scratch-your-own-itch'' model, to focus on thethesatisfaction and involvement of  end users~\cite{Terry2010}. Collection of the community feedback in OSS development is largely done by leveraging modern software engineering tools such as \textit{Issue tracking systems} (ITSs). ITSs allow OSS community members to create, discuss about, and manage the status of various system and project-related tasks, enhancements, problems, and questions, including those related to usability~\cite{Yusop2017}. Previous research demonstrated how the discussions on ITSs provide rich information to diverse community participants~\cite{arya2019analysis}. Particularly, usability issues are important venues to capture and record discussions about user experiences, opinions, desires, and justifications~\cite{cheng2018open}. However, because of the sheer amount of comments posted daily to ITSs, as well as the varied perspectives of different community members, contributors of OSS application projects face a major challenge in digesting the rich information embedded in ITSs in order to determine the actual user needs and consolidate the diverse feedback~\cite{Baysal2014}.


In this paper, we directly target this challenge through ArguLens (see Figure~\ref{fig:teaser}), a conceptual framework and machine learning-based technique leveraging an argumentation model to extract structured information from complex usability discussions. We formulate our research questions as follows:

\begin{enumerate}[noitemsep, leftmargin=0cm, label={}]
    \item \textbf{RQ1}: How do the open source software communities argue about usability issues in issue tracking systems?\vspace{4pt}
    \item \textbf{RQ2}: How effective are machine learning models in extracting arguments and their structure in usability issue discussions?\vspace{4pt}
    \item \textbf{RQ3}: To what extent can argumentation-enhanced representations of usability issue discussions support practitioners in understanding and consolidating community opinions and needs?
\end{enumerate}

To answer our research questions and construct ArguLens, we first conducted a content analysis on arguments in usability issue discussion threads of the Visual Studio Code GitHub project\footnote{https://github.com/microsoft/vscode}, using an adapted version of the Toulmin's argumentation model~\cite{toulmin2003uses}. Then we experimented with multiple supervised classifiers using different feature sets to extract argumentative information from the discussions. Finally, we performed a user study with participants who have experiences in using issue tracking systems to understand their perception of the argumentation-annotated usability issue discussion threads.

Our work produced the following outcomes and contributions. First, our content analysis of usability issue discussions resulted in an annotated corpus containing 5123 quotes (i.e. sentences or self-contained phrases) that serves as the foundation for ArguLens. Results indicated that contents in usability discussions broaden beyond the original post topic. In addition, we found that highly-disagreed arguments usually reside in the middle of the issue thread and thus can be easily overlooked. Second, we demonstrate that we can train supervised classifiers for automated argument identification, thus enabling argumentation-enhanced issue representation in ArguLens. Specifically, we observed that both Linear Support Vector Machine (SVM) and Naive Bayes classifiers provided satisfactory results using textual features based on term frequency-inverse document frequency (TF-IDF). Finally, our user study provided evidence that practitioners preferred representations of issue discussions supported by ArguLens. Removing non-argumentative contents, separating arguments by their topics, and providing information about argument components were appreciated by the participants when they focused on understanding the community's opinions in issue discussions. Our results also called for the needs of designing visual and interactive representations of the argument anatomy supported by ArguLens. In sum, ArguLens provides the direction of designing valuable tools and techniques for reasoning and discussing usability issues in a community-driven environment.

 
\section{Background and Related Work}
Our work is situated in the literature on (1) usability of open source projects, (2) argumentation theory and their applications to usability, and (3) automated argument mining.

\subsection{Open Source Usability}
Due to the distinctive characteristics of OSS projects, addressing their usability is particularly challenging. The usability of OSS applications is known to be negatively affected by factors such as developers' scarce awareness of users' needs, insufficient expertise, and an excessive emphasis on features and complexity rather than ease-of-use in OSS projects~\cite{Nichols2003}.
Along with the recent development of software engineering techniques and tools, as well as the increased awareness of usability in the software industry, OSS usability had become a growing research direction carrying important practical implications. The \textit{Issue Tracking Systems} (ITS) represent a common and fundamental tool for gathering the community members' needs and feedback about usability-related issues, as identified in several empirical studies about OSS usability practices~\cite{Bach2009, Nichols2006, Iivari2011}. The OSS community members (including users, developers, and other stakeholders) use ITSs to collaboratively raise, discuss, negotiate, and address usability-related topics. 

Although ITSs are useful in providing a common platform to engage the community, they suffer from several limitations when used for discussing usability issues in practice. For example, there is currently limited theoretical and practical support in classifying usability defects and handling the multifaceted usability discussions~\cite{Twidale2005, Yusop2017}. Current tools also do not accommodate the different communication styles adopted in diverse OSS communities~\cite{cheng2018open}. Moreover, community members engaged in ITS usability discussions usually experience an information overload, which hinders their effective participation and contribution~\cite{Baysal2014}. Because of these limitations, getting involved in OSS usability discussions is still a daunting affair, deterring OSS developers as well as other crucial stakeholders such as end users and user experience experts~\cite{Bach2009a, Rajanen2015, Iivari2013}. In this paper, we tackle this challenging problem of supporting OSS usability through the creation and evaluation of ArguLens to support OSS stakeholders in understanding and consolidating the diverse community opinions.

\subsection{Argumentation Theory and Its Application to Usability}
The study of argumentation has a long history due to the prevalent importance of persuasive speaking~\cite{herrick2017history}. Many researchers studied the structure of arguments, thus producing rhetoric and argumentation theory~\cite{toulmin2003uses, freeman1994toward}. Others applied these models to assist tasks in areas such as knowledge representation, legal reasoning, and negotiation~\cite{atkinson2006computational, bentahar2003commitment, bentahar2004computational, dung1995acceptability, mackenzie1979question}. 

Macewan et al's defined the term ``argumentation'' to emphasize its progressive property: ``argumentation is the process of proving or disproving a proposition. Its purpose is to induce a new belief, to establish truth or combat error in the mind of another''~\cite{macewan1898essentials}. We adopt this definition of \textit{argumentation} and additionally use the term \textit{argument} to refer to a concrete instance (e.g. a paragraph or a series of persuasive statements on a certain topic) produced via argumentation. 

Collectively, researchers have proposed many argumentation models for different purposes~\cite{bentahar2010taxonomy}. In order to understand how each argument towards usability unfolds in issue discussions, we adopt one of the most influential models, i.e., Toulmin's model of argumentation. Toulmin describes arguments using the metaphor of an organism, suggesting that an argument should have both a gross, ``anatomical'' structure and a finer, ``physiological'' one~\cite{toulmin2003uses}. To study the pattern of arguments, he identified six fundamental argumentative components:

\begin{itemize}[noitemsep, leftmargin=1.7cm]
    \item[\textbf{Claim}:] Conclusions or viewpoints one tries to convince others to agree with.
    \item[\textbf{Ground}:] Facts used to play as a foundation for the claim.
    \item[\textbf{Warrant}:] General or hypothetical statements, acting as bridges between data and claim.
    \item[\textbf{Qualifier}:] Indicating the strength granted by the warrant to support its claim.
    \item[\textbf{Rebuttal}:] Conditions in which the warrant is not applicable. As a result of a rebuttal, the consequential conclusion can be overturned.
    \item[\textbf{Backing}:] Demonstrating the validity of warrant.
\end{itemize}

\begin{figure}[b]
\centering
\includegraphics[width=0.9\linewidth]{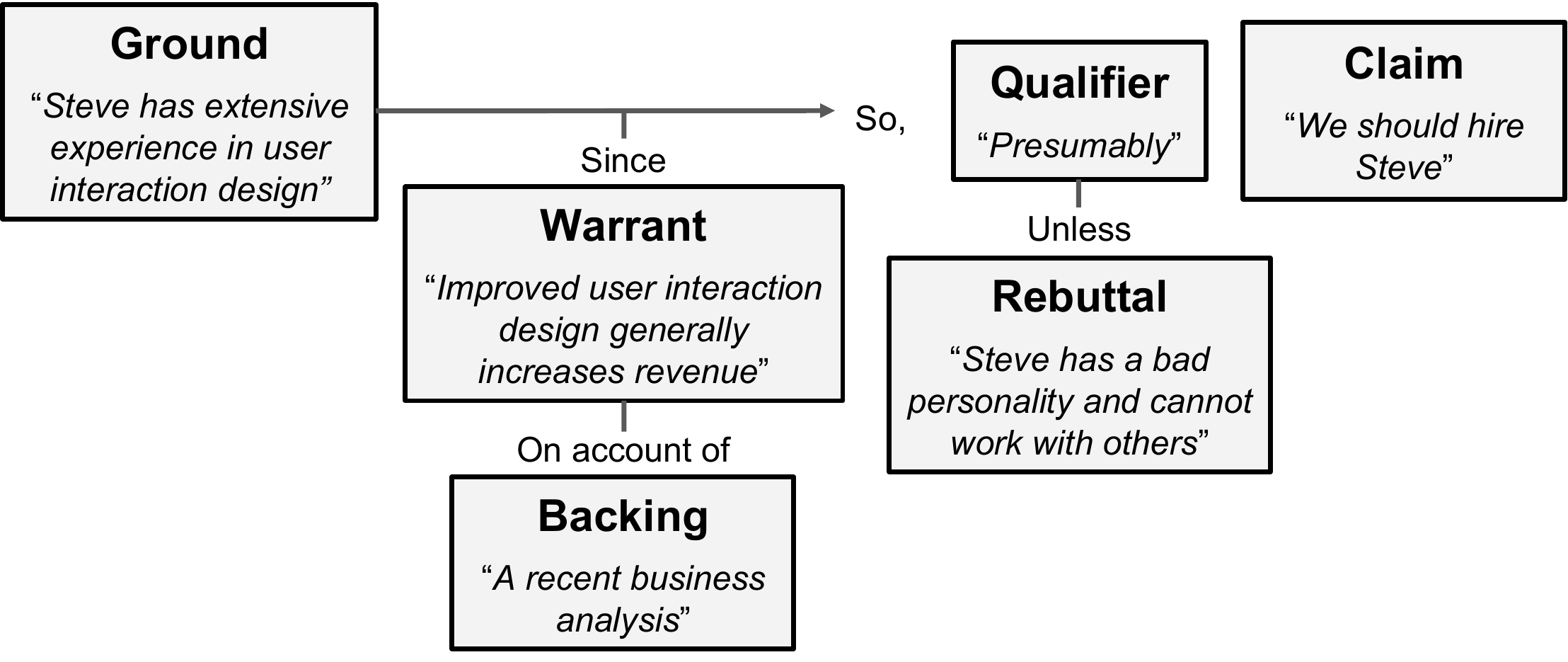}
\caption{Relationship among the six components of the Toulmin's model of argumentation~\protect\cite{toulmin2003uses} with examples.} 
\label{fig:toulmin_model_redraw}
\vspace{-4pt}
\end{figure}

Toulmin further incorporated their inter-relationships to form the layout of arguments, as shown in Figure~\ref{fig:toulmin_model_redraw}. 
 
Researchers have adopted Toulmin's argumentation theory for usability evaluations. For example, Nogaard et al. proposed that usability feedback can be viewed as an argument for a series of usability problems~\cite{norgaard2008evaluating}. By applying concepts from Toulmin's model and Aristotle's modes of persuasion, they provided guidelines about how to create persuasive usability feedbacks. Usability problems and solutions are also typically consolidated with a group of stakeholders involved in the design process~\cite{law2008consolidating}. A study with ten novice usability evaluators illustrated that their ability to defend their arguments during negotiation impacts feature/fix prioritization, which may further affect the design decisions~\cite{law2008consolidating}. Therefore, it is important to analyze argumentation in usability-related discussions. Our study contributes to this body of literature by adopting and adjusting Toulmin's model of argumentation to examine arguments contained in OSS usability issue discussion threads.


\subsection{Argument Mining}
Argument mining is the task of identifying argumentative contents and components in natural language texts \cite{sardianos2015argument}. Detecting argument components can help the readers consolidate ideas and support the argumentation authors to create a persuasive reasoning \cite{norgaard2008evaluating}. Arguments are extensively studied in formal written communications, such as legal documents and scientific publications~\cite{mochales2011argumentation, feng2011classifying, stab2014annotating, stab2017parsing}. Sentences in these documents are usually well-structured, which can facilitate the task of argument extraction. Comparing to formal sources, arguments taking place on online platforms are often less structured, vague, implicit, or simply poorly worded~\cite{boltuvzic2014back}. Several recent studies have focused on addressing these significant challenges in user-generated online texts. For example, Abbott et al. identified that contextual and dialogic features help in recognizing disagreement in informal political arguments~\cite{abbott2011can}. 
Florou et al. experimented using verbal tense and mood as features for the task of argument extraction of Web contents~\cite{florou2013argument}. Some researchers also worked on argument extraction from news, blogs and social media~\cite{goudas2014argument, sardianos2015argument}. 

Recently, Habernal et al. adapted Toulmin's argumentation model to Web discourse. Using an annotated dataset of 340 documents, they investigated the suitability of machine learning in automated identification of argument components. Their findings suggested that argumentation mining in user-generated Web discourse is a challenging but feasible task~\cite{habernal2017argumentation}. In this paper, we investigate the effectiveness of supervised machine learning techniques to detect arguments, along with argumentative components, in usability issue discussions. 
\section{Argumentation Analysis}
\label{sec:coding_study}
To build the foundation for ArguLens and answer RQ1 (\textit{How do the OSS communities argue about usability?}), we conducted a content analysis based on an argumentation model.

\subsection{Methods}
We carried out three main steps for argumentation analysis: (1) selecting OSS projects and issues, (2) adapting Toulmin's argumentation model, and (3) conducting the content analysis.

\subsubsection{Project and Issue Selection}
We focused our study on analyzing issue thread discussions for the Visual Studio Code\footnote{https://github.com/microsoft/vscode} project (VS Code), a source-code editor supported by Microsoft, published under the permissive MIT license. The reasons we selected this project are as follows. First, the application has a sophisticated graphical user interface which allows our study to focus on usability for end users. Second, the project is under active development and involves a large, heterogeneous community. Since the creation of the repository, an average of 353 issues are created each week. There are 906 recognized contributors, who made wide-ranging contributions including developing features, discussing issues, fixing bugs, conducting code reviews, and updating project documentation and websites. Lastly, the authors of this study are all experienced using this application. This ensures us to conduct a comprehensive and accurate analysis of the issue discussions in the project.

The issue selection was conducted in July 2018. Using GitHub REST API \cite{github_api}, we first ordered the closed issue threads by the number of comments. We then examined the title and the issue description for usability-relevance until five issues were selected. We only included closed issues in the analysis because they allow us to understand the complete flow of discussion until the issue is resolved. 
Table~\ref{tab:issuethreadcorpus} summarizes information about each issue we gathered.

\begin{table}[ht]
\centering
\small
\begin{tabular}{p{0.8cm} p{4.6cm} l}
\hline
\textbf{IssueID} & \textbf{Issue Title} & \textbf{\# of Comments} \\  \hline
224 & Proper tabs for open files & 411 \\
\clgrey{396} & \clgrey{Add support for opening multiple project folders in same window} & \clgrey{380} \\
4865 & Enhanced Scrollbar (add minimap) & 105 \\
\clgrey{9388} & \clgrey{Provide a setting so that only double click opens a file in the editor or expands a folder} & \clgrey{87} \\
14909 & Support a grid layout for editors & 191 \\
\hline
\end{tabular}
\caption{Information about issues in our corpus.}
\label{tab:issuethreadcorpus}
\vspace{-6pt}
\end{table}

\subsubsection{Argumentation Model Adaption}
In this study, we lay emphasis on examining the internal structure (i.e. anatomy) of usability-related arguments. While we acknowledge that some recent studies on argument mining (e.g.~\cite{Eger2017,stab2017parsing}) apply a simple claim-premise model, we decided to adopt a full model of argumentation (particularly Toulmin's model~\cite{toulmin2003uses}, described in related work). Our choice is grounded on our extensive experience working with OSS projects. Indeed, we believe that it is crucial for OSS contributors to identify (1) facts about the system (which can be captured by "ground" in the Toulmin model), and (2) general opinions about why a fact is relevant (corresponding to "warrant"). While simply distinguishing between `claim' and `premise' may result in a better accuracy in automated classification, we believe that such distinction is of limited usefulness in supporting the practical needs of OSS contributors that aim at understanding and prioritizing usability issues.

Toulmin's model focuses on documents containing well-formed explicit arguments. To adapt it to the informal arguments made in usability issue discussions, we started by directly using Toulmin's model to code two issues. Based on our observations, we made the following adjustments to better reflect the characteristics of ITS discussions.

First, in our initial analysis, we found that the boundary of \textit{Warrant} and \textit{Backing} is extremely difficult to draw in issue discussions. To avoid ambiguity, we merged those two components and annotated any statements acting as the bridge between user requests and concrete facts with \textit{Warrant}. Second, to adapt to the short, informal style of arguments in issue discussions, we removed \textit{Rebuttal} as an argument component and instead identified the standpoints of any comment in the discussion as \textit{Support} or \textit{Against} with respect to the original claim. Finally, similar to Habernal et al.~\cite{habernal2017argumentation}, we excluded \textit{Qualifier} because of its low representation in our issue discussion data. These changes resulted in the \textit{Components} coding schema used in the content analysis (see~Figure~\ref{fig:code_diagram}).

\subsubsection{Content Analysis}
\begin{figure}[t]
\centering
\includegraphics[width=0.9\linewidth]{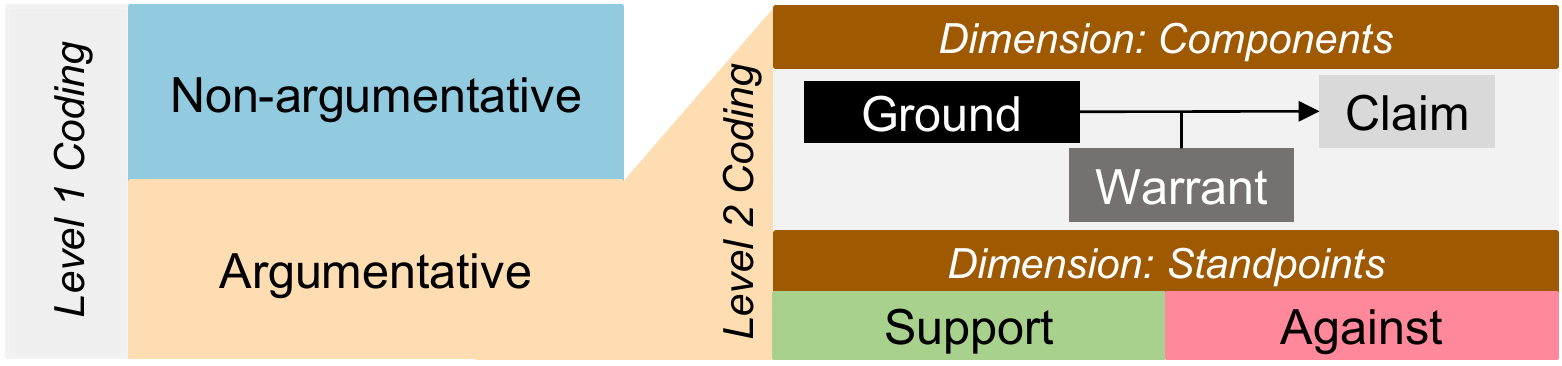}
\caption{Summary of the two levels of codes (including those in the two dimensions in the Level~2 coding) used in the content analysis process.}
\label{fig:code_diagram}
\vspace{-12pt}
\end{figure}

We performed the content analysis iteratively on two levels as illustrated in Figure~\ref{fig:code_diagram}. Upon data collection, we split the comments in each issue into individual sentences. In Level~1 coding, we coded each sentence as either \textit{argumentative}, which denotes that the sentence had a persuasive connotation, or \textit{non-argumentative}, which include all non-persuasive sentences. In Level~2 coding, we identified the characteristics of all the argumentative sentences based on two independent dimensions: (1) the argument \textit{components} based on the modified Toulmin's model (i.e. \textit{Claim}, \textit{Warrant}, or \textit{Ground}) and (2) the \textit{standpoint} of the author (i.e. \textit{Support} or \textit{Against}). During the coding process, if one sentence contains multiple argumentation components (e.g. claim + warrant), we manually split the sentence into segments to assign each segment a single, unambiguous code. We refer these sentences and self-contained segments as \textit{quotes}. Thus, all argumentative quotes were annotated using exactly one code from each of the aforementioned dimensions.

Furthermore, because an extended issue discussion usually contains multiple topics~\cite{arya2019analysis, cheng2018open}, we allowed coding multiple arguments in one issue thread to capture the diverse concerns people raised. We define an \textit{argument} as all the quotes that discuss the same claim, which may agree or disagree with the claim. A new argument can be brought up by making a claim about a new topic. Then, all discussions around this new claim are considered to be part of the new argument. We gave each argument a distinctive ID based on the chronological order of their first appearance in the issue discussion thread.

The first author conducted the initial coding for model refinement, with the support of two other authors when ambiguities or questions raised. Once the initial coding was finished, three researchers conducted six two-hour-long meetings and reviewed about 60\% of the quotes, including the most representative and questionable examples. The discussion was iterative, through which the team reached an agreement on all the discussed quotes and created a detailed codebook (see https://github.com/HCDLab/ArguLens). 
The first author finally used the codebook to re-code the non-discussed quotes.

To further assess the quality of our coding schema, we evaluated inter-rater reliability on a random sample of 200 non-discussed quotes. Four authors other than the first author each independently labeled 50 quotes using the codebook. The Coehn's~$\kappa$~\cite{Cohen:1960} against the label provided by the first author is 0.85 for Level 1 coding, 0.79 for argumentation components, and 0.80 for standpoints, indicating a ``substantial'' to ``almost perfect'' agreement~\cite{viera:garret:2005}.


\subsection{Results}
\begin{figure}[t]
\centering
\includegraphics[width=1\linewidth]{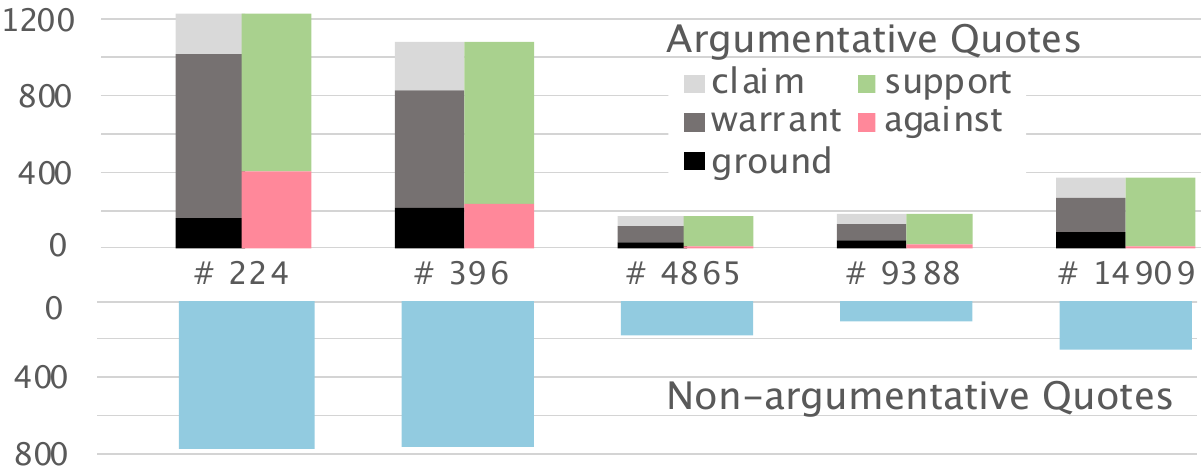}
\caption{Distribution of quotes coded with each of the adapted codes.}
\label{fig:codestatistics}
\vspace{-6pt}
\end{figure}

The content analysis generated a total of 5123 quotes from the five issue threads with a median of 621 quotes per issue ($IQR=1849-319.5$). 
Figure \ref{fig:codestatistics} shows the number of quotes annotated with each code for each issue.

\subsubsection{Level 1 Coding}
The first-level coding focused on identifying the argumentative comments in the issue threads and included two codes: 

\noindent\textit{Argumentative} (3034 quotes): Argumentative comments are made with the purpose of persuasion and fulfill a role of argument component (i.e., claim, ground, or warrant) in an issue discussion. These comments are the focus of our study and are further distinguished in Level~2 coding. 
As an example, Ivalexa made the following comment in issue 9388 to request implementing a desired behavior: ``\textit{Is it possible to add feature in order to force vscode to open file only by double clicking on it,  single click should only select a file in explorer?}''

\noindent\textit{Non-argumentative} (2089 quotes): This category includes any comments that are not persuasive in nature. We identified 11 concrete sub-categories of non-argumentative quotes such as questions (e.g., \textit{``Do you expect this behavior ... everywhere else ... or you just want it for the explorer?''}), workaround suggestions (e.g., \textit{``when I tried that earlier today it opened the file that received focus in the explorer tree immediately''}), progress inquiries (e.g., \textit{``Any news on this?''}), expressions of appreciation (e.g., \textit{``That's music to my ears!!''}), \textit{``+1''} or \textit{``-1''} alone in a sentence, etc. Please refer to the codebook in the auxiliary material for a complete list of non-argumentative sub-categories with examples.

\subsubsection{Level 2 Coding}
In this level, each argumentative quote is examined along two dimensions. The first dimension considers argument \textit{components} and contains the following three codes:

\noindent \textit{Component - Claim} (675 quotes): These quotes represent specific requests from the issue discussion participant. In our data about software usability in OSS, these quotes focused on describing users' expected behavior, proposing a new design, giving design suggestions, and expressing opinion towards a proposed design, etc. For example, in issue 4865, a participant provided the following design suggestion: ``\textit{I would just like you to consider showing selections and search results highlighted in the minimap as a feature.}''

\noindent \textit{Component - Ground} (541 quotes): These quotes illustrate facts about the current system or about a competitor' system reported by the discussion participant. There are two observed sub-themes under the \textit{Ground} code: (1) Report on the behavior of an existing system; for example: ``\textit{ctrl+tab between working files appears to work in most-recent order, as opposed to tabs which usually operate in the order they appear or are arranged.  (I've seen some editors support both).}'' (2) Description on how they usually interact with an existing system; for example: ``\textit{The first thing I ever do with VS Code is swap its close file and close active editor keyboard shortcuts, so that ctrl+W actually closes a file when I press it, rather than still leaving it in working files and cluttering up ctrl+tab.}''. 


\noindent \textit{Component - Warrant} (1818 quotes): This group of quotes is focused on connecting a \textit{Claim} and a \textit{Ground} with justifications such as potential use cases if the feature were implemented, the importance of following basic design principles, and personal feelings from using the existing systems. Most frequently, a \textit{Warrant} manifests use cases to demonstrate how the proposed changes (a \textit{Claim}) would fix the users' self-reported problem (i.e., a \textit{Ground}). For example, the following comment from issue 396 includes a description of a potential use case for the feature of ``opening multiple project folders in the same window'': ``\textit{As a Go developer, I find this feature extremely useful... I need to be able to quickly navigate to [many third-party libraries and projects] and read that code.}''


The second dimension of Level 2 coding models the \textit{standpoint} of the discussion participant and includes the following codes:

\textit{Standpoint - Support} (2347 quotes): This code includes comments showing agreement with the standpoint of the original argument. For example, in issue 224 an OSS participant supported adding proper tabs, commenting: ``\textit{Regardless of whether one can get by / used to `ctrl+tab' as an alternative, new VS code users will probably be put off by the lack of tabs.}''

\textit{Standpoint - Against} (687 quotes) This group of comments shows disagreement with respect to a particular argument. For example, opposing the argument given in the previous quote, another participant commented: ``\textit{I do not think tabs are a good way to show the list of open files unless you manage these things actively and close them.}''

\subsubsection{A Synthesized Example}
The following segment of discussion from issue 224 illustrates how our proposed framework can be applied in the context of OSS usability issues. In this example, the assigned codes from both the argument \textit{components} and the \textit{standpoint} dimensions are shown in pairs in square brackets at the end of each quote.

\begin{dialogue}
\speak{TurkeyMan} I really miss proper tabs for open files (like VS proper), and the ability to rip a tab out into its own window. \direct{warrant, support}\vspace{-6pt}
\speak{Nicte} +1 \direct{non-argumentative}. Tabs are the default way to work even in Visual Studio, to not mention other text editors such as sublime \direct{ground, support}.\vspace{-6pt}
\speak{nvivo}
I vote for adding tabs too.
\direct{claim, support}\vspace{-6pt}
\speak{RussBaz}
To be honest, one of the defining features of VS Code for me was an absence of tabs. \direct{warrant, against} I didn't even realise (at first) that they were missing! \direct{warrant, against} It was a very interesting point of view. \direct{warrant, against}\vspace{-6pt}
\speak{felixfbecker}
\lips Back when I used Atom, I hated it that \direct{warrant, against} every time I just quickly looked at a file, it would open a new tab, and within minutes my tab bar was full of useless tabs. \direct{ground, against} \lips
\end{dialogue}

\subsubsection{Multiple Arguments}
Extensive discussion of an issue can start from debating the needs for a requested feature and develop into considering design alternatives, more fine-grained design details, and opinions towards competitors' solutions. The following example (issue 224) shows a case where an OSS participant explicitly leads the discussion to the design of \textit{tabs} after a long discussion around the need for tabs: ``\textit{It's the top requested feature with thousands of votes. ... This conversation is pointless and needs to move on to ``How exactly should tabs work?''}

We identified a median of 24.0 ($IQR=49.0-15.5$) arguments per issue among the five issue discussion threads. The median number of discussion quotes per argument is 5.0 ($IQR=11.5-2.0$). Over all argumentative quotes in an issue thread, the median percentage of quotes belonging to arguments other than the original argument addressed in the issue post is 54.5\% ($IQR=62.1\%-32.4\%$).





\subsection{Discussion}
The content analysis results unveil several key characteristics of the arguments in usability issue discussions. First, non-argumentative comments constitute a large portion of the discussion. These comments usually include contents that are less relevant to the community's needs. Simply hiding them could help developers retrieve useful information and, therefore, would be a big step forward towards addressing information overload in a community-driven environment. Second, \textit{warrant} represents the majority component in the argumentative comments, followed by \textit{claim} then \textit{ground}. Furthermore, \textit{warrant}s usually include speculated use cases and personal opinions. These evidence echos with the previous observations that OSS discussions about usability are usually influenced by the personal opinions and experiences~\cite{Bach2009,cheng2018open}. Third, while all the issues were initially created for requesting specific changes in the existing interaction design, the discussion did not bound to the topic of the original post. Conversely, their content develops and broadens to include many related topics.

To better understand the characteristics of multiple arguments, we analyzed the \textit{Against} to \textit{Support} ratio among all arguments in the issue discussion threads to see how disagreed or controversial they were. We found that although the arguments from the original posts were the mostly-discussed in all five issues, they are not necessarily a highly-disagreed issue (their average \textit{Against}-\textit{Support} ratio is 0.115, comparing to the overall average of 0.325). We also observed that the highly-disagreed arguments (i.e. \textit{Against}-\textit{Support} ratio closer to $1$) usually reside in the middle of an issue thread, and are not necessarily heavily-discussed. One potential explanation is that these arguments may be hard to recognize in a lengthy thread. The current issue discussion platforms only display an issue thread in chronological order; this may cause the multiple requests and their corresponding discussion in the middle of the discussion threads to be easily overlooked.

\section{Machine Learning-Based Argument Extraction}
Annotated issue discussion threads for argumentation are essential for building intelligent tools supporting high-level reasoning tasks for the issue reporter, triager, assignees, and various stakeholders, including argument navigation, retrieval and trade-off analysis towards usability. However, manually labeling the comments is time-consuming, especially when the discussions become lengthy and the issue repository gets large. In this section, we explore the potential of using automated techniques in ArguLens to reduce the annotation effort and address RQ2: \textit{How effective are machine learning models in extracting arguments and their structure?}   

We used the 5123 quotes manually labeled in the previous section as the gold standard. While it would be ideal to adopt unsupervised methods to eliminate any manual effort on annotation, previous studies showed the limited performance on text-based unsupervised methods (e.g. topic modeling and clustering) and their sensitivity to parameter setting~\cite{agrawal2018wrong, arya2019analysis, chen2016survey}. Therefore, we experimented with supervised learning techniques to address the argument extraction task.  

\subsection{Methods}
\begin{figure}[t]
\centering
\includegraphics[width=1\linewidth]{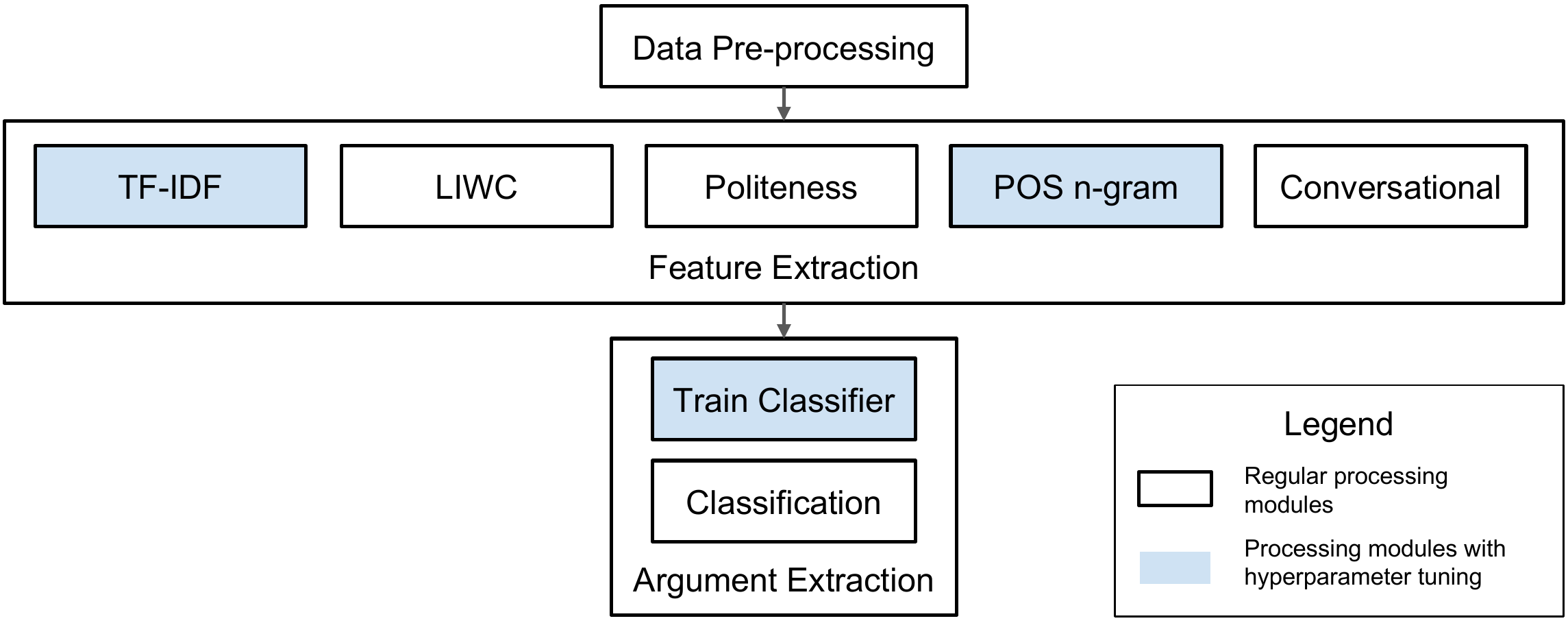}
\caption{Major modules of the supervised argument extraction process.}
\label{fig:machine_learning_process}
\end{figure}

Similar to the two-level coding analysis, the complete argument extraction is comprised of two-layer classification steps. Given one sentence from the issue discussion, the first layer classifies the sentence into either \textit{Argumentative} or \textit{Non-argumentative}. In the second layer, two separate classifiers label each argumentative sentence with topics from both dimensions of argumentation components (\textit{Claim, Warrant or Ground}) and standpoint (\textit{Support or Against}). 
The complete process includes three main steps: data pre-processing, feature extraction, and argument extraction (see Figure~\ref{fig:machine_learning_process}).

\subsubsection{Data Pre-processing}
We first cleaned up the dataset by removing the quotes that contain only non-alphabet characters and replaced the special contents with uniquely assigned tokens (see Table~\ref{tab:preprocess_special_token}). 
We then tokenized and lowercased each quote into individual words. Additionally, when extracting TF-IDF features and POS tag features (see the next section), we excluded punctuation marks and common contractions, and performed lemmatization on each word to obtain its dictionary form; for the other features, such as Politeness, we skipped this step because the information such as punctuation can be a strong indicator. We explain all the features and how we obtained them in the next section.

\begin{table}[!t]
\def\arraystretch{1.1}
\centering
\small
\begin{tabular}{l p{5cm}}
\hline
\multicolumn{1}{c}{\textbf{Special Token}} & \multicolumn{1}{c}{\textbf{Content Replaced}} \\ \hline
CODE\_BLOCK & multi-line source code blocks\\
\clgrey{CODE\_SEGMENT} & \clgrey{inline source code snippets} \\
QUOTE & quotations made in previous comments  \\
\clgrey{URL} & \clgrey{reference links to external resources}  \\
SCREEN\_NAME & mentions to GitHub users  \\
\clgrey{VERSION\_NUM} & \clgrey{version numbers} \\
PLUS\_ONE & token \textit{+} followed by numbers (i.e. +1) \\
\clgrey{MINUS\_ONE} & \clgrey{token \textit{-} followed by numbers (i.e. -1)} \\
ISSUE\_REFERENCE & token \textit{\#} followed by numbers (i.e. \#224) \\
\hline
\end{tabular}
\caption{Special tokens used during pre-precessing.}
\label{tab:preprocess_special_token}
\vspace{-12pt}
\end{table}


\subsubsection{Feature Extraction}
\label{subsec:feature_extraction}
We extracted two major types of features, i.e. Textual Features and Conversational Features, to train our supervised classifiers. Previous work on issue content analysis and argumentation mining~\cite{arya2019analysis, habernal2017argumentation} has demonstrated the effectiveness of using those features on related tasks with similar datasets.

\textbf{Textual Features} are extracted from the textual content of individual quotes. The fundamental assumption is that people potentially describe the same type of content in the argument using similar expressions, i.e., the words they use and the underlying social and physiological meanings they convey~\cite{tausczik2010psychological}. The concrete textual features used in our method include:

\begin{enumerate}[noitemsep, labelwidth=!, labelindent=0pt]
    \item \textit{Term Frequency-Inverse Document Frequency (TF-IDF):} a numerical measure that indicates how important a token is to a document in a corpus \cite{salton1988term}. Each word in corpus after data pre-processing acts as one feature weighted using the frequencies of words multiplied by their inverse document-frequency. We added n-grams as additional features to represent $n$ token sequences.
    \item \textit{Linguistic Inquiry and Word Count (LIWC):} a series of features obtained from LIWC, a text analysis program that captures a wide range of linguistic and psychological characteristics~\cite{pennebaker2001linguistic}. Previous studies have shown its effectiveness in various text analysis tasks, including detecting emotions, opinions, and bias on blogs and social media~\cite{tausczik2010psychological, yano2010shedding}.
    \item \textit{Politeness:} an overall measure ranging between 0 and 1 to reflect the key components of politeness theory, such as indirection, deference, personalization, and modality~\cite{danescu2013computational}.
    \item \textit{Part-of-Speech n-gram (POS n-gram):} a sequence of $n$ POS tags. POS tags indicate the part of speech, such as nouns, verbs, adjectives, etc. POS n-grams capture the structure of the sentence and therefore can potentially deduce subjective information comparing to simple text n-gram~\cite{bakliwal2011towards}.
\end{enumerate}

\textbf{Conversational Features} emphasize the characteristics of each quote within the discourse. It has been applied to processing software issues for issue thread summarization and information type detection~\cite{rastkar2010summarizing, arya2019analysis}. Similar to Arya et. al.~\cite{arya2019analysis}, our conversational features include five groups: (1) \textit{Participant features} describe the role that the comment author plays in the project (i.e. owner, collaborator, member, or other) and in the current issue thread (i.e. issue author or not); (2) \textit{Length features} depict the absolute length of a quote and its relative length with respect to other quotes in the issue comment and in the thread; (3) \textit{Structural Features} describe the location of the quote in the whole discussion threads; (4) \textit{Temporal Features} describe the time when the comment is made in relation to the immediately previous and next comment, as well as the whole discussion thread; and (5) \textit{Code Feature} indicates if the current comment contains code snippets.

\subsubsection{Argument Extraction}
Using the features extracted from issue quotes as input, we trained the machine learning classifiers for extracting arguments and their anatomy in the discussion quotes. Our candidate classifiers are Support Vector Machine (SVM) and Naive Bayes (NB) models. SVMs are well suited to deal with learning tasks where the number of features is large with respect to the number of training instances ~\cite{Joachims:1998,kotsiantis2007supervised}. In particular, we applied LinearSVM classifier with the class weight balancing technique to automatically adjust weights based on the class frequencies in the sample data. LinearSVM is computationally much more efficient than SVMs with non-linear kernels and yields state-of-the-art performance for high-dimensional sparse datasets, a typical scenario in text classification tasks like ours~\cite{Joachims:2006}. Performing a substantial feature selection would avoid dealing with such high dimensional input spaces. However, in supervised learning for text classification tasks, very few features are actually irrelevant, and feature selection often may cause a significant loss of information~\cite{Joachims:1998}. Thus, we did not perform any feature selection. 
We also considered Naive Bayes model because it tends to achieve good performance when the available training set is relatively small~\cite{kotsiantis2007supervised}. Specifically, we used the Complement Naive Bayes (CNB), which is an adaption of the standard Multinominal Naive Bayes (MNB) algorithm and is particularly suited for imbalanced data~\cite{rennie2003tackling}.

\subsubsection{Experiment Design}
We focused on three classification tasks: (Task 1) classifying Argumentative vs. Non-argumentative quotes, (Task 2) classifying the Claim, Warrant, and Ground components of argumentative quotes, and (Task 3) classifying Support vs. Against standpoints of argumentative quotes. We experimented building classifiers considering two independent variables:
\begin{enumerate}[noitemsep, leftmargin=0.6cm, label={({{\arabic*}}})]
    \item The machine learning model to adopt with two levels: LinearSVM and Complement NB.
    \item Feature set to use as input for the machine learning models with six levels: TFIDF, LIWC, Politeness, POS n-gram, Conversational features, and concatenating all five aforementioned feature sets together. 
\end{enumerate}


The combination of the two independent variables resulted in 12 configurations. To evaluate the performance of all configurations, we adopted the standard nested stratified 5-fold cross-validation~\cite{friedman2001elements}, which contains two layers. The inner layer is used to tune the hyperparameters for certain modules in our techniques (modules that need hyperparameter tuning are shown with a blue shade in Figure~\ref{fig:machine_learning_process}); e.g., for Linear SVM, the tuned hyperparameter is the cost parameter C, which indicates the penalty of misclassification on the training set. The outer layer is then used to evaluate the performance of the models on unseen data. 
The performance of the models is evaluated using average Precision, Recall, and F1-Measure from the outer layer; these metrics are commonly used to evaluate classification tasks~\cite{powers2011evaluation}. For a certain class in question, Precision measures how accurate the classifiers' predictions on the class are correct; Recall measures how effective the classifier is in predicting the correct labels for all instances from that class; F1-Score is the harmonic mean of the Precision and Recall. We used python and the Scikit Learn library to implement the experiments and calculate all metrics. 


\subsection{Results and Discussion}
\begin{figure}[t]
\centering
\subfloat[\small{Average F1 measures of each configuration for Task 1} 
]{
\label{fig:model_performance_argumentative}
\includegraphics[width=0.99\linewidth]{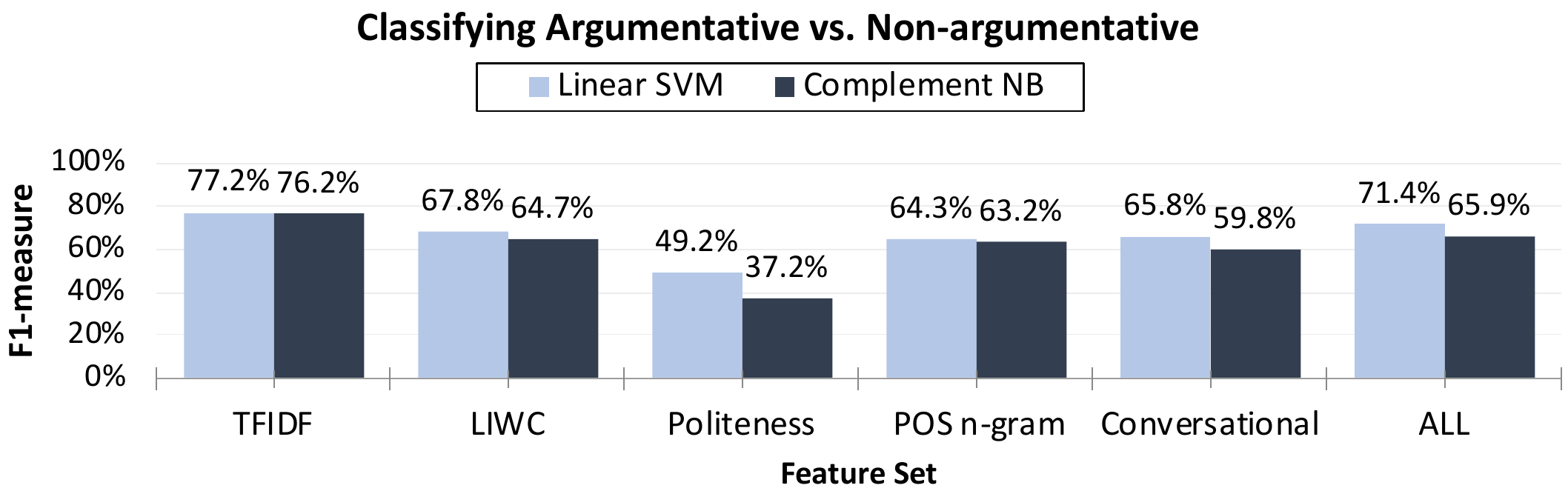}}\hfill
\subfloat[\small{Average F1 measures of each configuration for Task 2} 
]{
\label{fig:model_performance_rebuttal}
\includegraphics[width=0.99\linewidth]{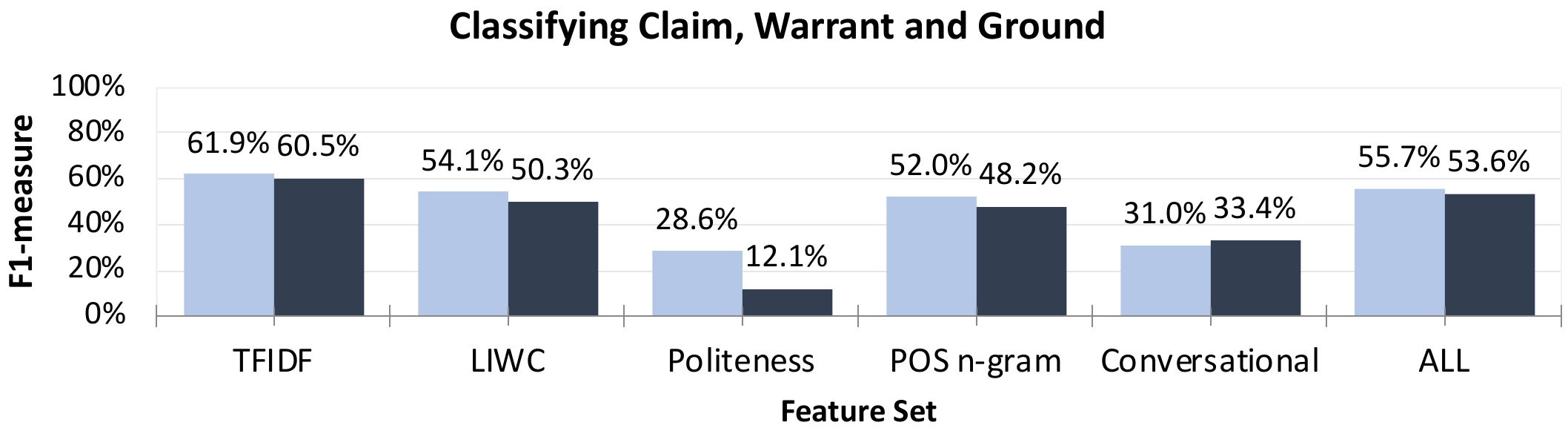}} \hfill
\subfloat[\small{Average F1 measures of each configuration for Task 3} 
]{
\label{fig:model_performance_argu_part}
\includegraphics[width=0.99\linewidth]{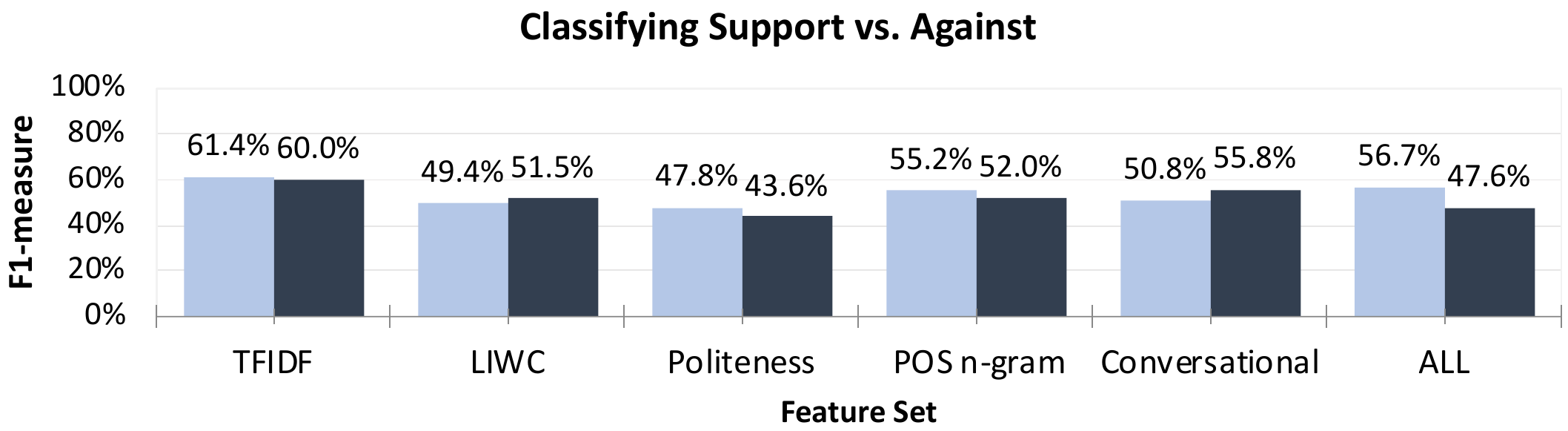}}
\caption{Model Performance for each classification task.}
\label{fig:model_performance}
\vspace{-10pt}
\end{figure}

Figure \ref{fig:model_performance} summarizes the average F1-scores of the 12 configurations for the three classification tasks. Our results indicate that using TF-IDF features yielded the highest F1-score. One possible explanation is that our annotated corpus includes a large portion of domain-specific words. TF-IDF features excelled other feature sets at capturing domain knowledge from the text, which might be strong indicators for their argumentation roles. This result also has positive practical implications because TF-IDF features are relatively easy to obtain compared to other feature sets. On the other hand, the selection of different classifiers was not as critical as the feature sets used. In fact, given the same set of features, we observed a negligible difference in performance between Linear SVM and Complement Naive Bayes classifiers. 

The best configuration for all three classification tasks was obtained using Linear SVM with TF-IDF features, achieving an average F1-score of 77.22\%, 61.92\%, and 61.42\% for the three tasks respectively. We report the Precision, Recall and F1-score for all three tasks in Table~\ref{tab:automated_result_all}. As a baseline, we also report the performance of a trivial classifier always predicting the majority class (i.e., Argumentative, Warrant, and Support, for the three tasks, respectively). This is a consolidated practice in machine learning for text categorization as the baseline performance gives an indication of the inherent difficulty of the classification task itself~\cite{Sebastiani:2002,Yang:1999}.

\begin{table}[t]
\centering
\small
\begin{tabular}{l p{11mm}p{11mm}p{8mm}p{11mm} } 
 \hline
\textbf{Label to Classify} & \textbf{Precision} & \textbf{Recall} & \textbf{F1} & \textbf{Support} \\ \hline
Argumentative & 0.80 & 0.85 & 0.82 & 608 \\
Non-argumentative & 0.76 & 0.68 & 0.72 & 418\\ 
\clgrey{Task 1 Average/Total} & \clgrey{0.78} & \clgrey{0.77} & \clgrey{0.77} & \clgrey{1025} \\
\hdashline
Baseline & 0.29 & 0.50 & 0.37 & 1025 \\
\hline

Claim & 0.63 & 0.52 & 0.57 & 135 \\
Warrant & 0.74 & 0.83 & 0.78 & 394 \\
Ground & 0.56 & 0.46 & 0.50 & 109 \\
\clgrey{Task 2 Average/Total} & \clgrey{0.64} & \clgrey{0.60} & \clgrey{0.62} & \clgrey{608} \\
\hdashline
Baseline & 0.21 & 0.33 & 0.25 & 608 \\
\hline

Support & 0.83 & 0.84 & 0.83 & 470 \\
Against & 0.41 & 0.40 & 0.40 & 138 \\
\clgrey{Task 3 Average/Total} & \clgrey{0.62} & \clgrey{0.62} & \clgrey{0.61} & \clgrey{608} \\
\hdashline
Baseline & 0.39 & 0.50 & 0.44 & 608 \\
\hline

\end{tabular}
\caption{Detailed measurements for three classification tasks with configuration of Linear SVM and TF-IDF features, as well as the baseline (always predicting the majority class) performance. Support refers to the number of instances in the testing set for that label.}
\label{tab:automated_result_all}
\vspace{-6pt}
\end{table}

For all tasks, we observed substantial improvements over the baseline classifier. Specifically, the first level classification, i.e. Argumentative versus Non-argumentative, achieved the best result. Notably, the Precision and Recall for the Argumentative class are both more than 80\%. Such strong performance demonstrates the possibility of designing tools that leverage supervised classification models to remove a large number of irrelevant content for the issue readers. In terms of the argument components and standpoints, our classifiers performed better for classes that contain more data instances, i.e. Support, and Warrant. However, its lukewarm performance on Against, Claim, and Ground suggests that the user might need to accumulate more annotated training data to achieve a better result. While not perfect, our techniques can be used in a semi-automated way to suggest the underlying argument anatomy in the issue discussion, or be plugged into tools for empirical studies involving mining overall trends in ITSs.

As regard to the classification approach, we would like to remind the reader that our goal in this work is not to advance the latest NLP techniques for argument classification. Rather, we aim at demonstrating the potential of anatomizing usability issue discussions using an argumentation framework, thus paving the road for renovating how people can collaboratively address open source usability issues. We are aware of recent advancements in NLP research leveraging word or sentence embedding and neural architectures for text categorization. However, the practical value of such approaches has been questioned in many software engineering areas because of the difficulty to acquire a large amount of training data and to disambiguate domain-specific terms~\cite{Fakhoury2018,Fu2017}. Meanwhile, we hope that our method can serve as a strong baseline to attract more researchers to apply advanced argumentation-mining techniques on this challenging and important problem.


\section{Practitioner Evaluation}
To better understand the ArguLens approach from the perspectives of experienced OSS practitioners (RQ3), we performed a within-subject user study with eight frequent users of GitHub Issues. This study is approved by the Institutional Review Boards of Polytechnique Montreal and McGill University.

\begin{figure*}[t]
\centering
\includegraphics[width=\linewidth]{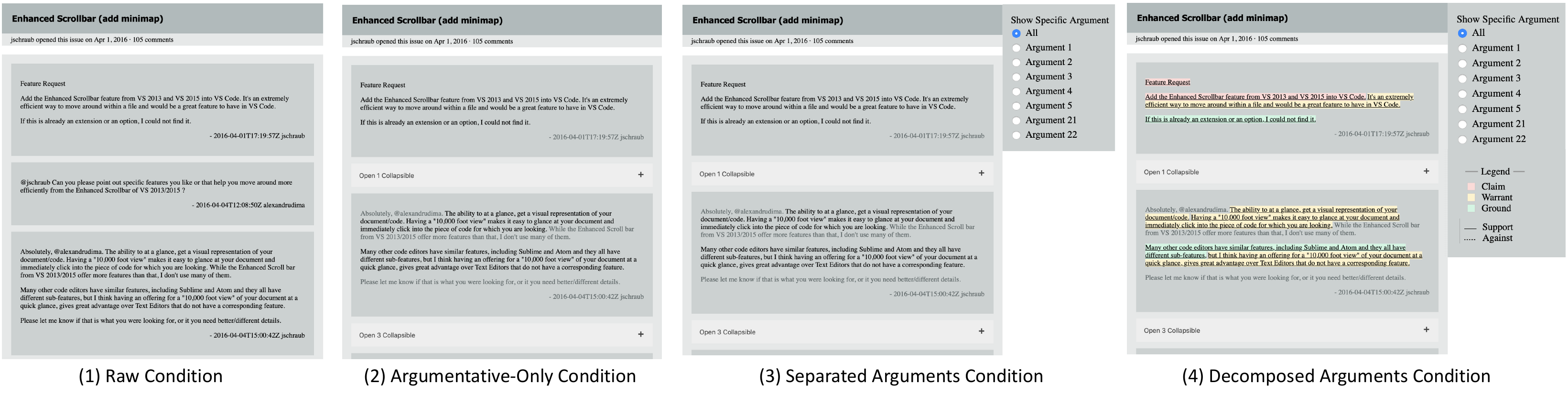}
\caption{Stimuli used in the user studies with OSS practitioners (using issue 4865 as an example).}
\label{fig:stimuli}
\vspace{-6pt}
\end{figure*}

\subsection{Methods}
We recruited eight participants (three females) from personal contacts. All participants had experience with GitHub and its \textit{Issues} feature, having contributed to at least one project on the platform.
They all had used VSCode as a text or code editor; some used it on a daily basis at the time of the study.

Each user study took about 1.5 hours to complete. During the study, we first conducted a semi-structured interview with each participant to collect demographic information. We then asked participants to review four issue discussion threads that we analyzed in the previous steps; to make the review a manageable task within the context and the time frame of the user study, we trimmed each issue discussion to only include the first 50 comments. We developed web-based stimuli (see Figure~\ref{fig:stimuli}) to represent the issues in four increasingly complex conditions. Consistently with our intention of exploring practitioners' perspectives about the argument anatomy itself, we focused on presenting an interface with minimal interaction design. All but the first condition leveraged the ArguLens technique and all annotations about arguments, argumentation components, and standpoints were based on the manual content analysis results. The four conditions are:

\begin{enumerate}[noitemsep, leftmargin=12pt]
    \item \textbf{Raw}: Participants see the original discussion comments with no processing.
    \item \textbf{Argument-Only}: Participants see the discussion comments with non-argumentative sentences greyed out or put into collapsed sections (if the entire comment is non-argumentative).
    \item \textbf{Separated Arguments}: Participants see the argumentative comments (same as Condition~2) and a list of radio buttons as a sidebar indicating different arguments in the thread. The participants can further navigate among the arguments by clicking on the radio buttons, making all other arguments greyed out or put into collapsed sections.
    \item \textbf{Decomposed Arguments}: Participants see the same content and have the same argument filtering ability as Condition~3. However, all argumentative sentences they see are color-coded and highlighted according to their argumentation components and standpoints.
\end{enumerate}

We followed a within-subject experimental design; the order of the conditions and the condition-issue assignments were counterbalanced using Zeelenberg and Pecher's approach~\cite{Zeelenberg2015}. At the beginning of each issue review, participants were provided with a one-page description about the concepts related to the corresponding condition and could ask any clarification questions. While reviewing each issue, the participants were asked to (1) describe the main topics discussed, (2) identify the potential positive and negative impacts on users when the issue is fixed,
and (3) decide whether the issue should be fixed. We also encouraged the participants to think-aloud during these tasks. At the end of each issue review, the participants completed a System Usability Scale (SUS) questionnaire~\cite{brooke1996sus} and a Subjective Mental Effort Question (SMEQ)~\cite{Sauro2009Questionnaires}. The SUS questionnaire is commonly used to quantify the perceived usability of a system; higher SUS scores (ranging from 0 to 100) indicate better usability; a meta review of 2,324 studies indicated a mean SUS score of 70 across various systems~\cite{Bangor2008SUS}. SMEQ includes one rating question (ranging from 0 to 150) for perceived mental effort; lower scores indicate a lower mental effort. Once all issues were reviewed, we asked the participants to select the most and the least useful representation and to provide rationales for their choice.

\subsection{Results and Discussion}
All but one participant completed the four tasks; the remaining participant (P2) did not evaluate the \textit{Separated Arguments} representation due to time constraints. The average SUS and SMEQ scores are summarized in Table~\ref{tab:SUS_SMEQ}. Given the qualitative nature of the study, in the rest of this section we focus on reporting the themes emerged from participants' feedback.

\begin{table}[h]
\centering
\small
\begin{tabular}{lllll}
\toprule
 & Raw & Arg-only & Sep. Args. & Dec. Args. \\
\midrule
SUS & 78.21 & 84.38 & 87.86 & 71.88\\
SMEQ & 24.50 & 18.13 & 17.86 & 34.25 \\
\bottomrule
\end{tabular}
\caption{Average SUS and SMEQ scores for the four conditions.}
\label{tab:SUS_SMEQ}
\end{table}

Our results indicated that designing the representation of the anatomy of arguments is an important topic. Among our eight participants, seven appreciated having the possibility to hide or gray out the non-argumentative comments for the purpose of determining the community opinions when reviewing the issues. For example, P6 mentioned: ``\textit{It [hiding the non-argumentative sentences] already helps a lot to quickly get the idea about the issue and what people are thinking about it.}'' P5 also mentioned that simply highlighting the argumentative comments and sentences helped ``\textit{differentiate things that you want to put more emphasis on.}''. Only one (P3) preferred the \textit{Raw} representation, because it presents an interface familiar to the participant and includes all contextual information; P3 commented: ``\textit{I don't like having just the arguments. I would like to read a little bit more about the background.}'' 

Four participants considered the \textit{Separated Arguments} representation as the most useful. They thought that separating arguments by their topics helped them to be more focused when considering the diverse opinions. For example, P5 mentioned ``\textit{I think that feature [being able to quickly switch among the arguments] alone is really great. ... it's really helpful to be specific.}'' P1 also thought that having separated arguments is ``\textit{useful to summarize long issues and get main points quickly.}''

Two participants found the \textit{Decomposed Arguments} the most useful. They both used the argument components and standpoints extensively while reviewing the issues to determine the main topics and the potential impacts to users. For example, P8 said: \textit{``The first question [main topic of the issue] was probably the claim here ... the support and against could determine the other two [positive and negative impacts]... and probably, if I count how many support and against I have in the whole issue, I can say that it's a good idea to implement it or not.''}

Although the \textit{Decomposed Arguments} provides useful supports for issue review, four participants disliked this representation because of its conceptual and visual complexity. For example, P3 explained: ``\textit{I can say that having three colors in here and underline and everything... like, it makes reading hard.}'' Participants also indicated that understanding this representation requires some learning. For example, P6 said: ``\textit{The colors are not very intuitive in a way, but it may be that I'm just not used to it.}'' P2 also commented: ``\textit{I think this would be very powerful ... if someone has been using this for a longer time.}''

\section{General Discussion}
ArguLens has the potential to support a wide range of applications and at the same time has limitations that can be addressed in the future. We discuss these topics here.

\subsection{Potential Applications}
ArgueLens can be directly used for tools that visualize the argument anatomy of community-generated usability discussions. Our user study showed that such anatomy helps the participants focus on the most relevant information when reviewing usability issues. Effective interaction and visual design leveraging ArgueLens would then provide the much-needed support for OSS stakeholders (e.g. developers, designers, and end users) to better comprehend and consolidate community opinions about the usability issues of the system.


Moreover, ArgueLens introduces an effective, yet straightforward framework of usability argumentation. Tools that encourage or reinforce such framework can support OSS communities to move towards a shared convention and a common language of discussing usability-related topics, which in turn supports community engagement in such challenging issues.

Leveraging ArguLens, new automated or semi-automated techniques can also be developed to support analytic activities on OSS usability issues at the project level. For example, discussion summary and prioritization tools of usability issues can be developed based on ArguLens. Such efforts would then further bolster OSS developers' empathy towards users, thus encouraging them to better incorporate the end users' voices.

\subsection{Limitations and Future Work}
Due to the effort required for the task of manual coding, we performed an in-depth analysis on only five usability issues from one OSS project. This might have introduced a threat to external validity that we mitigated by focusing on the most heavily discussed issues that involved diverse community members of an active OSS project. In this way, we were able to address the complex situations where support for understanding the community opinions is most needed. Nevertheless, the ArguLens framework and the performance of the machine learning techniques can be affected by the unique characteristics of the analyzed issues and the project. Future work extending this analysis to other usability issues and projects would allow establishing its external validity. Additionally, our work focused on the usability issues because of their increasing importance in OSS applications and the overwhelming challenges involved to address them. Although we speculate that ArguLens can be applicable to other types of OSS issues prone to extended community discussions (e.g. privacy or security), as well as other types of community-driven platforms, these applications require future work.

Further, although our machine learning solutions performed well in certain tasks (e.g. discriminating Argumentative and Non-argumentative sentences), the performance was less desirable in others. Particularly, its performance is sensitive to the amount of the training data, indicating that extended manual labeling effort is still necessary to achieve its utility in practice. Furthermore, given the current precision and recall achieved by the argument extraction tasks, human-in-the-loop method is necessary to support the ArguLens technique. Future work might also explore combining supervised and unsupervised approaches to further reduce the manual effort. 

Finally, while our user study focused on a typical application scenario of reviewing a usability issue to determine its potential impact on the community, the study certainly does not cover all the scenarios potentially supported by ArguLens. Moreover, the stimuli used in our user study provided only a basic presentation of the embedded information and did not focus on offering a sophisticated visual and interaction design. Thus, exploring an effective design of tools that leverage the ArguLens techniques is an important future work.


\section{Conclusion}
We proposed and evaluated ArguLens, a conceptual framework and machine learning-based technique that leverages an argumentation model to support comprehension and consolidation of OSS community's opinions on usability issues. We created a corpus of 5123 quotes, annotated with an argumentation framework as the foundation for ArguLens. Leveraging supervised machine learning on this corpus, we evaluated various argument extraction techniques; our best performing classifier achieved satisfying results, paving the way to the implementation of semi-automated approaches and tools. Our study with experienced ITS users revealed that ArguLens can help practitioners get focused and retrieve the relevant information when reviewing issue discussions. In sum, as an important step forward, ArguLens can accelerate building effective methods and tools that aim to improve OSS usability.

\section{Acknowledgement}
We thank our participants in the user study for their thoughtful efforts. This work was funded by the Natural Sciences and Engineering Research Council of Canada (NSERC) [RGPIN-2018-04470, RGPIN-2019-05403].

\balance{}

\bibliographystyle{SIGCHI-Reference-Format}
\bibliography{OSSUsability,CherylThesis}

\end{document}